# Circular Antenna Array Design for Breast Cancer Detection

K. Ouerghi[(1)], N. Fadlallah[(2)], A. Smida[(1)], R. Ghayoula[(1)], J. Fattahi[(3)] and N. Boulejfen[(4)]

[(1)] Unit of Research in High Frequency Electronic Circuits and Systems, Faculty of Mathematical, Physical and Natural Sciences of Tunis, Tunis El Manar University, Tunisia
[(2)] RADIOCOM, Institut of Technology –saida P.O.Box 813 #36 Lebanon
[(3)] Department of Computer Science and Software Engineering. Laval University, Quebec, Canada.
[(4)] Research Center for Microelectronics and Nanotechnology, Technopole Sousse, 4050, Sousse, Tunisia

*Abstract*— —Microwave imaging for breast cancer detection is based on the contrast in the electrical properties of healthy fatty breast tissues. This paper presents an industrial, scientific and medical (ISM) bands comparative study of five microstrip patch antennas for microwave imaging at a frequency of 2.45 GHz. The choice of one antenna is made for an antenna array composed of 8 antennas for a microwave breast imaging system. Each antenna element is arranged in a circular configuration so that it can be directly faced to the breast phantom for better tumor detection. This choice is made by putting each antenna alone on the Breast skin to study the electric field, magnetic fields and current density in the healthy tissue of the breast phantom designed and simulated in Ansoft High Frequency Simulation Software (HFSS).

## I. INTRODUCTION

Breast cancer is the most prevalent type of cancer among females in the globe [1]-[2], which is generally a fast cell growth within the breast tissue[3], usually in the epithelium of the lobules and ducts[4]. It can also circulate to other parts of the human body if it is a metastatic breast cancer. These cancer cells can propagate to other parts of the body, such as liver, lungs, bones and brain. The cancer cells divide up and become taller out of control, thus forming new tumors. Even though the new tumors are getting bigger in another part of the body, it is still always a breast cancer[5]. Since the last medical examination of breast cancer in 2008, cases of this type of cancer are rising by more than 20%, with 14 % of mortal cases [1]-[6].In 2012, 1.7 million women were diagnosed with breast cancer all over the world [6]. This tumor is now considered as the most commonly diagnosed tumors among women and an inevitable concomitant of death with 522 000 death cases. To downsize the risks of breast cancer mortality, we should diagnose the appearance of the malignant tissue in healthy tissues and treat this serious illness in its early stages.

To quickly identify the presence of breast tumor, we are going to proceed with the study of microwave breast imaging (MBI).This technique has been widely used especially for biomedical diagnosis[7]-[8].

The principle behind the MBI consists of using a transmitter microwave signal to emit signals inward the breast and receivers to detect those emitted signals after they interact with the breast. In the presence of a tumor , usually with higher dielectric properties [9]than those of the other tissues of the breast, the amount of signal energy scattered by the tumor is higher than the one scattered by the fabrics of a normal breast with no tumor. A model of breast close to the real human breast should be modeled using the Ansys HFSS software with a semi-spherical geometric shape. The different dielectric properties of the mammary tissues that make up the breast model should be close to the real breast properties[10].

In this paper, we present the design of an inset-fed rectangular patch antenna for microwave imaging using a 2.45GHz signal. Also, we investigate the performance characteristics of five antennas working in the same frequency range by placing them on the breast skin to obtain an antenna that satisfies the design criteria for 3-D antenna array system for microwave breast imaging. This investigation is conducted by replacing the geometry of patch with slots and geometry of ground while maintaining the same resonant frequency.

## II. MICROWAVE SYSTEM

### A. Antenna design

The antenna is a key element in the microwave imaging system. To perfectly detect the tumor, an adapted antenna is required. Our antenna design strategy starts with a basic rectangular inset-fed micro-strip patch antenna resonating at 2.45GHz with a total dimension of 37.26x28.82mm on an FR-4 substrate with a relative permittivity ($\varepsilon_r$)of 4.4, a width of 65.4mm, a length of 88.99mm, and a thickness of 1.588mm, as shown in Fig.1.(1) and Fig.3.(a). Five different studding antennas are placed on the skin of the breast shape to investigate the different values of the electric as well as magnetic fields and the current density of a healthy breast tissue with and without a tumor inside the breast shape to eventually define the best antenna giving the perfect response. Fig1 present the evolution of our first antenna from a not editing antenna to four other modified antenna structures. Starting with the second antenna (2) on the same figure, a modification at the level of the ground dimensions is done with -65.4x48.82mm strating from the up of patch to the end of the microstip line . the third and fourth (3), (4) antenna we kept the modified ground and we added 2 different slot in the middle of the patch from the right and the left position. The fifth antenna have kept the first basic groud with 65?4x88.99mm dimention and sloting the ground, Position of the slot is determined to work out at inset distances .

Hence, the antennas are modeled, simulated and optimized in the Ansoft HFSS software. Then, the different obtained results are evaluated to reach the part of breast shape design.

specifically in the microwave frequency. In this paper, however, we adopted a hemispherical shape to model the breast phantom composed of a skin with an outer radius of 70 mm and a thickness of 2mm, a Fatty tissue named healthy tissue with a 68-mm outer radius and a spherical tumor placed in the middle of the breast with a 10-mm radius, as shown in Fig. 3 (b).

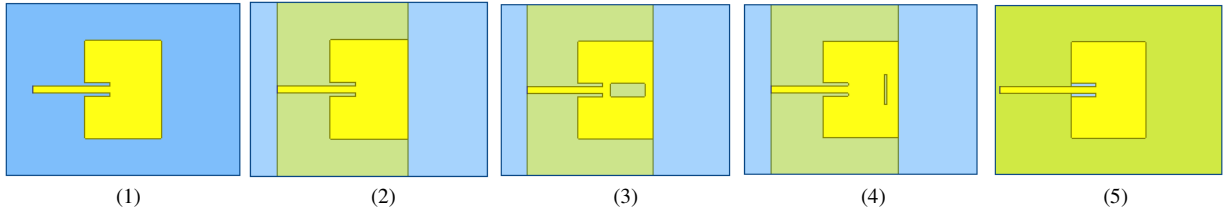

(1)　　　(2)　　　(3)　　　(4)　　　(5)

Figure. 1. used antennas structure from 1 to 5

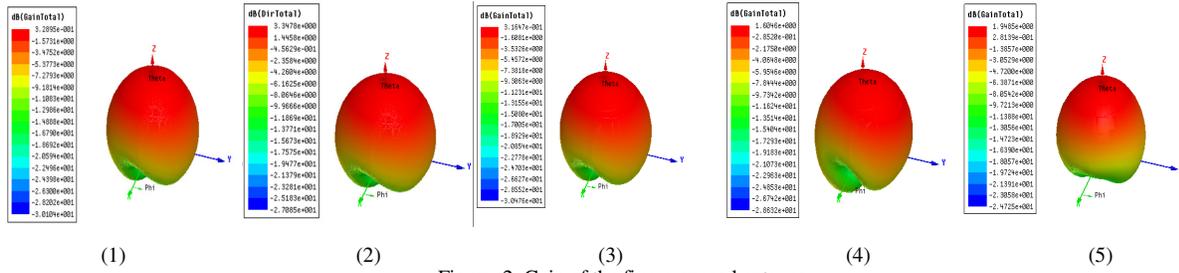

(1)　　　(2)　　　(3)　　　(4)　　　(5)

Figure. 2. Gain of the five proposed antenna

TABLE I. ELECTRICAL PROPERTY OF BREAST TISSUE

| Breast tissue | Dielectric | Conductivity(S/m) |
|---|---|---|
| Healthy tissue | 36 | 4 |
| Skin | 9 | 0.4 |
| Tumor | 50 | 4 |

TABLE II. ANTENNA PARAMETER USED

| Antenna parameter | Value (mm) | Antenna parameter | Value (mm) |
|---|---|---|---|
| W | 65.4 | L1 | 3.997 |
| L | 88.99 | L2 | 13.84 |
| Wp | 37.26 | GL | 9.57 |
| Lp | 28.82 | GW | 1 |
| LG | 48.82 | FL | 20 |
| W1 | 4 | FW | 3.036 |
| W2 | 11.26 | ~ | ~ |

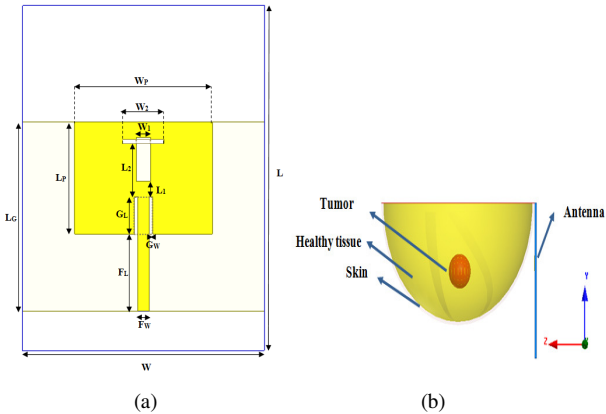

(a)　　　(b)

Figure 3. (a) Different parameter used in the five studied antennas (b) antenna on breast model

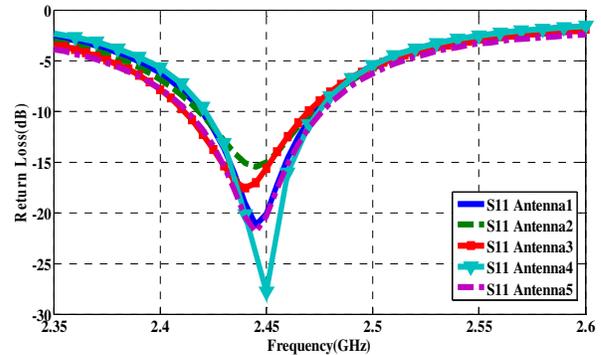

Figure. 4. S11 of the five proposed antennas

### B. Breast phontom design

Different designs of breast phantoms have been used by researchers [11]-[12]-[13]-[14] All these phantoms are characterized by essential electrical properties which are the relative permittivity $\varepsilon_r$ and conductivity$\sigma$[4]-[15]. An extended research has proved that there is an important contrariety between healthy and malignant breast tissue properties, The different electrical properties of the breast and tumor are presented in Table II.

Fig. 4 shows s the simulation results of the s-parameters in accordance with the frequency of the five studied antennas.

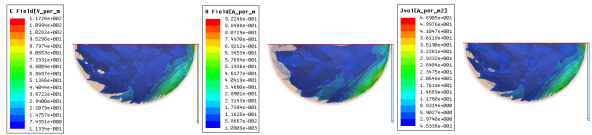
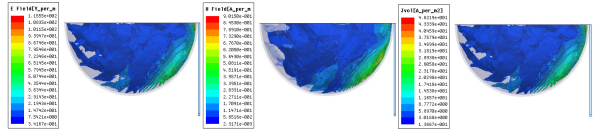

Antenna1

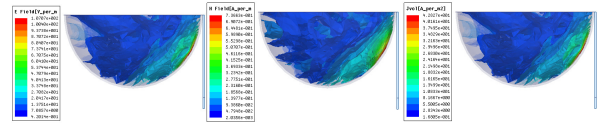
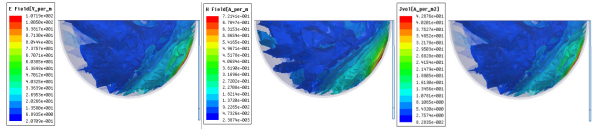

Antenna 2

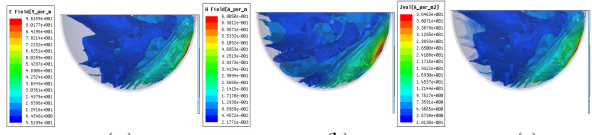
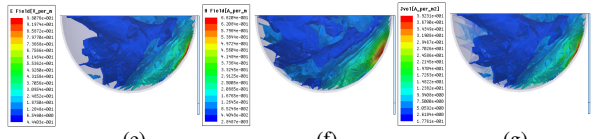

Antenna 3

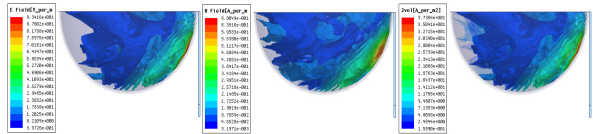
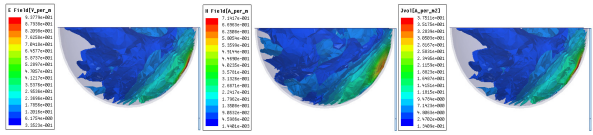

Antenna 4

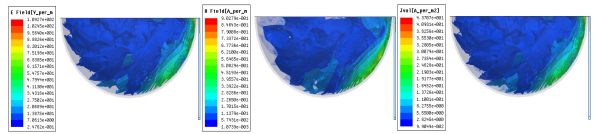
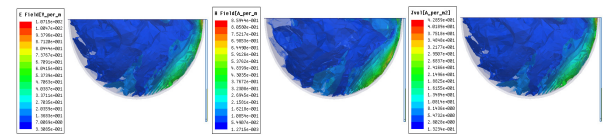

Antenna 5

Figure. 5. (a) Electric field E at breast tissue level without tumor (b) Magnetic field H at the breast tissue level without tumor (c) Current density J at breast tissue level without tumor (e) Current density J at breast tissue level with tumor (f) Magnetic field H at the breast tissue level tissue with tumor (g) Current density J at breast tissue level tissue with tumor

TABLE. III. RESULTS OBTAINED FROM THE ELECTRIC AND MAGNETIC FIELDS, AND THE CURRENT DENSITY FOR EACH ANTENNA ELECTRICAL PROPERTY OF BREAST TISSUE

| Antenna number | Electric Field (Vm) | | Magnetic Field (Am) | | Courant Density (AM²) | |
|---|---|---|---|---|---|---|
| | Without Tumor | With Tumor | Without Tumor | With Tumor | Without Tumor | With Tumor |
| **Antenna1** | 115.55 | 117.26 | 0.901 | 0.922 | 46.21 | 46.90 |
| **Antenna2** | 107.19 | 107.07 | 0.721 | 0.736 | 42.87 | 42.82 |
| **Antenna3** | 98.07 | 96.15 | 0.662 | 0.68 | 39.23 | 38.46 |
| **Antenna4** | 93.77 | 93.41 | 0.714 | 0.68 | 37.51 | 37.36 |
| **Antenna5** | 107.15 | 109.27 | 0.859 | 0.902 | 42.85 | 43.707 |

The adaptation of our antennas at the same frequency of 2.45GHz, as shown in Fig.4 This adaptation is done with different levels of S-parameters, namely -21.23dB for antenna(1), -15.50dB for antenna(2), -17.59dB for antenna (3), -27.81dB for antenna (4) and -21.88 dB for antenna (5).. The radiation patterns vary from 3.34dB to 1.6dB for the fives antennas, as shown in Fig. 2.c. The signals reflected by the breast tissue can reveal if there is a tumor. We compare the results of the breast phantom model with and without tumor exposed in the near field of antennas radiating a permissible 0.034 mW at 2.45GHz.to notice the difference between the electric fields, magnetic fields and current density of the healthy tissue in both cases. Fig.5 shows the different results desired to study the five antennas going from 1 to 5 where each one is placed on the breast model without tumor in (a), (b) and (c) and with tumor in (e), (f) and (g). To better understand these results, Table. III reformulates s them and clarifies the difference between the results of the breast tissue with tumor and those of the breast tissue without tumor in antennas 1, 2, 3, 4 and 5 , which cannot help distinguish the nature of tumor.

Compared with the other antennas, it can be seen that, there is a decrease in the values of the E-field, H-field and the current density of 109.27Vm, 0.902Am and 43.707Am², respectively for antenna (5) with the presence of the tumor ; and of 107.15Vm, 0.859Am and 42.85Am², respectively for antenna (5) without the presence of tumor. Therefore we represent specific absorption rate (SAR) of the fifth antenna in the skin and fatty tissue of the breast with no tumor and with tumor , The SAR value refers to power averaged over 10 g of tissue. results are shown in Fig6, It is visible that for a healthy breast the current SAR values over the skin and fatty tissue are lower. This study shows the proposed antenna can be used to effectively detect tumor inside breasts.

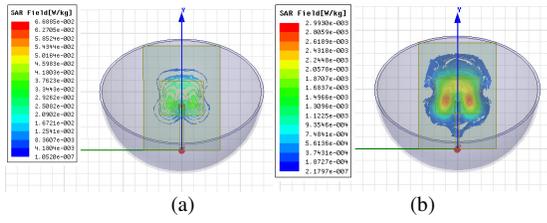

(a)                  (b)

Figure. 6. SAR distribution on breast phantom without tumor (a) skin tissue (b) FAtty tissue

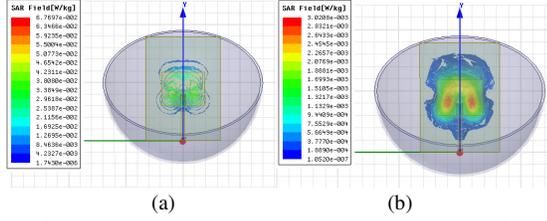

(a)                  (b)

Figure. 7. SAR distribution on breat phantom with tumor (a) skin tissue (b)fatty tissue

This study is proofvof concept which needs to be verified experimentally. We believe with an array antenna the sensitivity for detection can be greatly enhanced.

### C. Antenna Array

In this section, we study the feasibility of a circular antenna array. To improve the performance of the microwave imaging system, we have chosen to put our antenna (antenna (5) already adopted in the previous section  after the comparative study allowing for the best result) in front of the simplified breast model  to  nearly end up in the form of a half sphere to permit  more antennas to be placed in the network. We have arranged the array in a circular configuration where 8 antennas are used nearly close to each other, connected to lumped ports and separated by a circularly distance of λ/2, as explained in Fig.8.

The simulation results of the antenna array shown in Fig.9 and Fig. 10 reveal that it successively performs the radiation pattern in the resonance frequency of 2.45GHz with a peak of 12.23 dB compared with the element antenna with 1.94 dB. The mutual coupling between antennas is illustrated in Fig.8. which shows that theyare nearby $S_{21}$, $S_{31}$, $S_{41}$, $S_{51}$, $S_{61}$, $S_{71}$ and $S_{81}$, and in the desired frequency their maximum part is below -20dB. Accordingly, it is clear that this antenna array is adequate for use in the microwave breast imaging (MBI). All the antennas are connected through wave guide ports of 50-ohm impedance. For all the radiators, a discreet port is attached to the feeding point  acting as a waveguide port. The complete antenna array se-tup in HFSS is shown in Fig. 8.

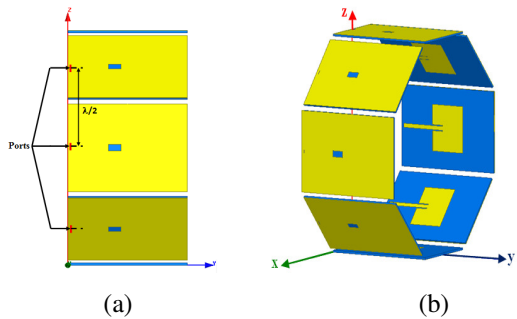

(a)                  (b)

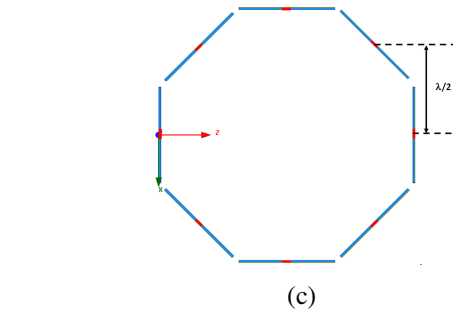

(c)

Figure. 8.circular antenna array (a)Side view(b) Perspective view (c) Top vue

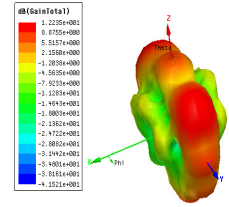

Figure. 9. radiation pattern of antenna array

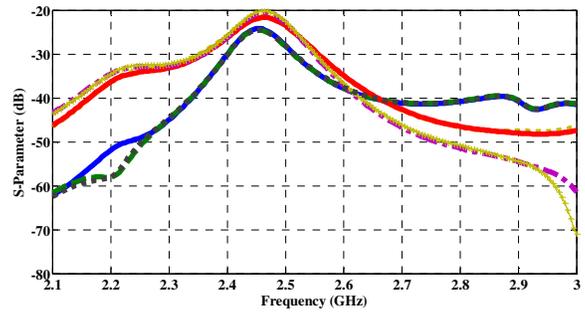

Figure. 10. Mutual coupling of the array

We can define the term mutual coupling as the situation when two or more neighbor antennas come close to each other. While one antenna transmits a part of energy, another receives it; and both exist in transmitting and receiving mode.Among the annular array configuration in transmitting mode, the antenna used to radiate a part of energy which is received by the other is known as mutual coupling. Mutual coupling decreases when the distance between neighbor antennas increases.

### III. CONCLUSION

In this paper, the studied antenna is intended for use in a scanning system for antenna array based on microwave breast cancer detection. This system  differs from the proposed antenna design  in several ways. A comparative study of five microstrip patch antennas is made to chose the perfect antenna that can be used in microwave breast imaging system to identify maliniant tissues developing in the women breast.

This study  proves that antenna (5)  guarantees a decrease in the results of the electric fields, magnetic fields and current density  inside  healthy tissue with the existance of malignat tumor in the breast compared with a breast without a tumor. Compared with other works, we have developed a simlified antenna array to enhance better imaging for tumor detection.The simulation results of the antenna array show a good impedance matching with low mutual coupling and high radiation pattern.This array proved to be efficient and easily made for Microwave Breast Imaging (MBI) compared with other imaging techniques. The antenna exhibited a good directional radiation pattern with acceptable gain.

NOTICE